\documentclass{emulateapj}

\def\saxj{SAX J1808.4--3658}
\def\igrj{IGR J00291+5934}

\def\Porb{P_{\rm orb}}

\def\T0{T^*_0}
\def\asini{a_1 \sin i}

\begin{document}

\title{Timing an Accreting Millisecond Pulsar: Measuring the Accretion 
Torque in \igrj}

\author{L. Burderi\altaffilmark{1}, T. Di Salvo\altaffilmark{2},
G. Lavagetto\altaffilmark{2}, M.T. Menna\altaffilmark{3},
A. Papitto\altaffilmark{3,4}, A. Riggio\altaffilmark{2},
R. Iaria\altaffilmark{2}, F. D'antona\altaffilmark{3},
N. R. Robba\altaffilmark{2}, L. Stella\altaffilmark{3}} 

\altaffiltext{1}{Universit\`a degli Studi di Cagliari, Dipartimento
di Fisica, SP Monserrato-Sestu, KM 0.7, 09042 Monserrato, Italy;  
email: burderi@mporzio.astro.it}
\altaffiltext{2}{Dipartimento di Scienze Fisiche ed Astronomiche,
Universit\`a di Palermo, via Archirafi 36 - 90123 Palermo, Italy}
\altaffiltext{3}{I.N.A.F. - Osservatorio Astronomico di Roma, via Frascati 33,
00040 Monteporzio Catone (Roma), Italy}
\altaffiltext{4}{Dipartimento di Fisica, Universit\'a degli Studi di Roma 
''Tor Vergata'', via della Ricerca Scientifica 1, 00133 Roma, Italy}

\begin{abstract}

We performed a timing analysis of the fastest accreting millisecond pulsar
\igrj\ using RXTE data taken during the outburst of December 2004. 
We corrected the arrival times of all the events for the orbital
(Doppler) effects and performed a timing analysis of the resulting phase
delays. In this way we have the possibility to study, for the first time 
in this class of sources, the spin-up of a millisecond pulsar 
as a consequence of accretion torques during the X-ray
outburst.  The accretion torque gives us for the first time an independent
estimate of the mass accretion rate onto the neutron star, which can be compared
with the observed X-ray luminosity.
We also report a revised value of the spin period of the pulsar.
\end{abstract}


\maketitle

\section{Introduction}
The so-called recycling scenario links two different classes of 
astronomical objects, namely the millisecond radio pulsars (usually
found in binary systems) and the Low Mass X-ray Binaries (hereafter LMXBs),
or, at least, a subgroup of them. The leading idea of this scenario
is the recycling process itself, during which an old, weakly 
magnetized, slowly spinning neutron star is accelerated by the accretion 
of matter and angular momentum from a (Keplerian) accretion disk down 
to spin periods in the millisecond range. In this way, at the end of 
the accretion phase, the neutron star rotates so fast that it is
resurrected from the radio pulsar graveyard, allowing the radio pulsar
phenomenon to occur again despite the weakness of the magnetic field.

Although this scenario was first proposed long ago (see
e.g. Bhattacharya \& van den Heuvel 1991 for a review), the most embarassing
problem was the absence of coherent pulsations in LMXBs. Only recently, the
long seeked millisecond coherent oscillations in LMXBs have been found,
thanks to the capabilities (the right combination of high temporal resolution
and large collecting area) of the RXTE satellite. In April 1998, a transient
LMXB, \saxj, was discovered to harbour a millisecond pulsar ($P_{\rm spin}
\simeq 2.5$~ms) in a compact ($P_{\rm orb} \simeq 2$~h) binary system
(Wijnands \& van der Klis 1998; Chakrabarty \& Morgan 1998).  We now know
seven accreting millisecond pulsars (Wijnands 2005; Morgan et al. 2005); all
of them are X-ray transients in very compact systems (orbital period between
40 min and 4 h), the fastest of which ($P_{\rm spin} \simeq 1.7$~ms), \igrj,
has been discovered in December 2004 (Galloway et al.\ 2005, hereafter G05).

Timing techniques applied to data of various accreting millisecond pulsars,
spanning the first few days of their outbursts, allowed an accurate
determination of their main orbital parameters.  However, only a few attempts
have been made to determine the spin period derivative (Chakrabarty et al.\
2003; Galloway et al.\ 2002). The first reported measurement of a spin-up 
in these sources was made for IGR J00291+5934 (Falanga et al.\ 2005, F05).

In this paper we apply an accurate timing technique to the fastest currently
known accreting millisecond pulsar, \igrj, with the aim of constraining the
predictions of different torque models with good quality experimental data.
Our results indicate quite clearly that a net spin up occurred during the
December 2004 outburst of \igrj\ (see also F05) and that 
the derived torque is in good agreement with that expected from matter 
accreting from a Keplerian disk.

\section{The Timing Technique}

For a periodic pulsating source, the time of arrival of a given pulse at
the solar system baricentre is affected by three effects which causes
temporal delays with respect to the predicted ones. These effects are 
1) uncertainties in the orbital parameters, 2) uncertainty in the spin 
frequency and possible secular variations, 3) uncertainties in the 
source position.  
In standard timing techniques (see e.g.\ Blandford \& Teukolsky 1976) the
predicted arrival time of a given pulse is computed using a first guess of the
parameters of the system, and the difference between the experimental and
predicted arrival times, namely the residuals, are fitted with a linear
multiple regression of the differential corrections to the parameters.  This
means that the differential correction to orbital parameters, spin frequency 
and its derivative, source position in the sky, are computed 
simultaneously. This technique has the
obvious advantage to give a self-consistent solution, where all the
correlations in the covariance matrix of the system are fully taken into
account. However, the convergence of the fit is not always guaranteed and --
especially in the case of long temporal baselines - convergence to secondary
minima might lead to inaccurate solutions. 

On the other hand, if the orbital period is much shorter than the timescale on
which the spin period derivative and the source position uncertainty are 
expected to produce a significant effect, it is easy to see that the delays 
in the arrival times produced by the uncertainties in the orbital parameters 
are distinguishable from those caused by the other two effects. 
This is because the former oscillate on the orbital period timescale, while
it is expected that the latter will follow a secular trend dictated by the
torques on the accreting neutron star and by the orbital motion of Earth.
In the following we therefore discuss all these effects separately and 
describe in general the timing analysis suitable for these cases.

The procedure we applied is the following.
In order to obtain the emission times, $t_{\rm em}$, the
arrival times of all the events, $t_{\rm arr}$, were firstly reported to
the Solar system barycenter adopting the best estimate of the source 
position in the sky. We note here that the contribution of the relative
acceleration of the binary system with respect to the Solar system could 
in principle produce not negligible contribution to the frequency derivative.
For a source located at the position of \igrj\ at a distance of few kpc, the 
most relevant of these effects is due to the planar acceleration along the 
Galactic plane, which gives a spin derivative 
$\dot \nu_{\rm Gal} \sim 5 \times 10^{-20}\; \nu$
(see e.g.\ Damour \& Taylor 1991). This is several orders of magnitude
lower than the spin frequency derivatives we found (see below), and therefore
in the following these kind of effects are not taken into account.
Then we corrected for the delays caused by the binary motion using the best 
estimate of the orbital parameters through the (first order approximated) 
formula:
\begin{equation}
\label{eq:corr}
t_{\rm em} \simeq t_{\rm arr} - x \left[ \sin ( m + \omega ) + \frac{e}{2}
\sin ( 2 m + \omega ) - \frac{3}{2}\; e \sin \omega \right]
\end{equation}
where $x = a \sin i /c$ is the projected semimajor axis in light seconds, 
$m = 2\pi (t_{\rm arr} - T^*)/ \Porb$ is the mean anomaly, 
$T^*$ is the time of ascending node passage at the beginning of the
observation, $\omega$ is the periastron angle, and $e$ the eccentricity. 
In the following, for simplicity, we use $t$ instead of $t_{\rm
em}$.  From eq.~(\ref{eq:corr}) we compute the observed phases as:
$\phi = \nu (t - T_0)$, where $\nu = 1/P_{\rm spin}$ and $T_0$ is the 
start time of the observation. On the other hand,
the expected phase variations, $\delta \phi_{\dot \nu}$, caused by a 
spin frequency derivative, $\dot \nu$, can be computed by a double direct 
integration:
\begin{eqnarray}
\label{eq:dotnudel}
\delta \phi_{\dot \nu}(t) = \int_{T_0}^{t} \left[ \int_{T_0}^{t^{'}}  
\dot \nu(t^{''}) \,
dt^{''} \right] dt^{'} .
\end{eqnarray}
In the simplest case of a constant $\dot \nu$, the integration gives
a parabolic function of time. In general,
fitting these expected phase variations to the observed ones, 
we can obtain an estimate of $\dot \nu (t)$ and hence important 
information on the torques acting on the accreting neutron star. 

To obtain these information is important to evaluate any source of
error in the observed phase variations; we start discussing the
errors induced by the uncertainties on the orbital parameters of the
binary system. The differential of $\phi = \nu (t - T_0)$, with $t$ 
given by expression~(\ref{eq:corr}), with
respect to the orbital parameters allows to calculate the uncertainties in the
phases, $\sigma_{\phi\,\rm orb}$, caused by the uncertainties, $\sigma$, in 
the estimates of the orbital parameters:
\begin{eqnarray}
\label{eq:orbdel}
\sigma_{\phi\,\rm orb} = \frac{x}{P_{\rm spin}} \left\{
\sin^2 m \left(\frac{\sigma_{x}}{x} \right)^2  +
\cos^2 m  \right. 
\nonumber \\
\left. \left[ m^2  
\left(\frac{\sigma_{P_{\rm orb}}}{P_{\rm orb}} \right)^2 
+ \left(\frac{2 \pi \sigma_{T^*}}{P_{\rm orb}} \right)^2 \right]
+ \sin^2 m \; \cos^2 m \; \sigma_{e}^2
\right\}^{1/2} 
\end{eqnarray}
The uncertainties in the adopted orbital parameters will result in a 
``timing noise'' of amplitude $\sigma_{\phi\,\rm orb}$. 
These should be therefore added in quadrature to the statistical 
uncertainties $\sigma_{\phi\,\rm stat}$ on the experimentally determined
phase delays. The resulting uncertainties on the phase delays
will be: $\sigma_{\phi} = (\sigma_{\phi\,\rm orb}^2 + 
\sigma_{\phi\,\rm stat}^2)^{1/2}$.

On the other hand, the uncertainties in the phase delays,
$\sigma_{\phi\,\rm pos}$, caused by the uncertainties on the estimates of 
the source position in the sky, will produce a sinusoidal oscillation on the 
Earth orbital period. For observation times shorter than one year, as it 
is the case for most transient accreting millisecond pulsars, this can 
cause systematic errors on the determination of the neutron star period
and its derivative, since a series expansion of a sinusoid contains a
linear and a quadratic term.
In order to evaluate these effects, let us consider the expression of the phase
delays induced by the Earth motion for a small displacement, $\delta \lambda$ and 
$\delta \beta$, in the position of the source in ecliptic coordinates, $\lambda$
and $\beta$ (see e.g.\ Lyne \& Graham-Smith 1990):
\begin{equation}
\label{eq:posdel}
\Delta \phi_{\rm pos} = \nu_0\, y \,\left[\,\sin (M_0 + \epsilon) \cos \beta\, 
\delta \lambda - \cos (M_0 + \epsilon) \sin \beta \,\delta \beta \,\right]
\end{equation}
where $y =  r_{\rm E}/c$ is the distance of the Earth with respect to the Solar 
system barycenter in light seconds, 
$M_0 = 2 \pi (T_0 - T_\gamma)/P_\oplus - \lambda$, where $T_\gamma$ is the 
time of passage through the Vernal point, and 
$\epsilon = 2 \pi (t - T_0)/P_\oplus << 1$ in our case.

After some algebraic manipulations, this can be written as:
$\Delta \phi_{\rm pos} = \nu_0\, y \,\sigma_\gamma \,\sin (M_0 + \epsilon - \theta^*)\; u$,
where $\sigma_\gamma$ is the positional error circle,
$\theta^* = \arctan (\tan \beta \; \delta \beta / \delta \lambda)$, and
$u = [(\cos \beta\; \delta \lambda)^2 + (\sin \beta\; \delta \beta)^2\; ]^{1/2}
/ \sigma_\gamma$.
Since the true source position must lie within the error circle, the
following inequalities hold: $(\cos \beta\; \delta \lambda / \sigma_\gamma)^2 \le 1$, 
$(\delta \beta / \sigma_\gamma)^2 \le 1$, and thus $u \le (1+\sin^2 \beta)^{1/2}$. 
This means that the uncertainty in the source position is: 
\begin{equation}
\label{eq:posdel2}
\Delta \phi_{\rm pos} \le \nu_0\, y \,\sigma_\gamma \,(1+\sin^2 \beta)^{1/2}\,
\sin (M_0 + \epsilon - \theta^*)
\end{equation}
We can expand it in series in the parameter $\epsilon << 1$ in order to find 
the systematic uncertainties induced on the linear and quadratic term. Note that,
since the values of $\delta \lambda$ and $\delta \beta$ are the differences between 
the nominal and the true (unknown) source position (which can be everywhere 
within the error circle), their ratio is undetermined, and hence 
$\theta^*$ can be any value between $0$ and $2 \pi$. 
We have therefore maximized 
the functions $\sin \theta^*$ and $\cos \theta^*$ with $1$ separately in the
linear and quadratic terms of the series expansion. The resulting systematic
error in the linear and the quadratic term of phase delays evolution versus time,
which correspond to the spin frequency correction and the spin frequency 
derivative, respectively, are: 
$\sigma_{\nu\, {\rm syst}} \le \nu_0\, y \,\sigma_\gamma \,(1+\sin^2 \beta)^{1/2}
\,2\,\pi\,/P_\oplus$
and $\sigma_{\dot \nu\, {\rm syst}} \le \nu_0\, y \,\sigma_\gamma \,
(1+\sin^2 \beta)^{1/2}\,(2\,\pi\,/P_\oplus)^2$.

Summarizing, the phase variations caused by $\dot \nu$ are effectively 
distinguishable from those induced by the uncertainties in the orbital
parameters, which result in a ``timing noise'' of amplitude given by 
eq.~(\ref{eq:orbdel}). On the other hand,
the uncertainty on the source position cannot be easily decoupled from
the phase variations caused by $\dot \nu$ (particularly for observation
times much shorter than one year), and therefore results in systematic
errors on the estimate of the spin frequency and its derivative.

\section{Observations and Data Analysis}

\igrj\ was observed by RXTE between 2004 December 3 and 21. While the
observations between December 3 and 6 were already analyzed in G05, in this
paper we analyze the data between December 7 and 21 taken from a public ToO.
We mainly use data from the RXTE Proportional Counter Array (PCA, Jahoda et
al.\ 1996), which consists of five identical gas-filled proportional counter
units (PCUs), with a total effective area of $\sim 6000$ cm$^2$, sensitive in
the energy range between 2 and 60 keV. We used data collected in generic
Events mode, with a time resolution of $125~\mu$s and 64 energy channels.
These files were processed and analyzed using the
FTOOLS v.5.3.1. In order to eliminate the Doppler effects caused by the Earth
and satellite motion, the arrival times of all the events were converted to
barycentric dynamical times at the Solar system barycenter. The position
adopted for the source was that of the proposed radio counterpart (which is
compatible with that of the proposed optical counterpart, see Rupen et al.\
2004; Fox \& Kulkarni 2004). For the spectral analysis we also used data
from the High-Energy X-ray Timing Experiment (HEXTE, Rothschild et al. 1998, 
$\sim 20-200$ keV energy range).

We corrected the arrival times of all the events for the delays caused by the
binary motion using eq.~(\ref{eq:corr}) with the orbital parameters given in
G05.  In order to check for the presence of the pulsations during our
observation, we performed an epoch folding search on each continuous interval
of data (lasting on average 60 min) around the spin period given in G05. The
pulsation was clearly visible up to December 12.  After that, owing
to poor statistics, pulsations could be detected only by folding $\sim 1$~day 
worth of data. No pulsations were detected after December 14, in accordance
to what reported in G05.

Using our longer temporal baseline (about 7 days) with respect to that in G05
(about 3 days), we firstly tried to increase the accuracy of the orbital period 
measurement using the technique described in Papitto et al.\ (2005) 
and successfully applied to \saxj. However, no significant improvement 
was found. 
Adopting the uncertainties in the estimates of the orbital parameters given in
G05 in eq.~(\ref{eq:orbdel}), 
we obtain: 
$\sigma_{\phi\, \rm orb} \la 0.01$, 
where we have maximized $\sin$ and $\cos$ functions with 1, and used 
$t-T_0 \la 7$ days.
Therefore, we expect that the uncertainties in the orbital parameters 
will cause a ``timing noise'' not greater than $\sigma_{\phi\,\rm orb} 
\times P_{\rm spin} \sim 0.02$~ms.

To compute phases of good statistical significance we epoch folded each
interval of data in which the pulsation was significantly detected at the spin
period given in G05 with respect to the same reference epoch, $T_0$,
corresponding to the beginning of our observations. The fractional part of the
phase was obtained fitting each pulse profile with a sinusoid of fixed
period. To compute the associated errors we combined the statistical errors
derived from the fit, $\sigma_{\phi\,\rm stat}$, with the errors
$\sigma_{\phi\, \rm orb}$	 
as $\sigma_{\phi}=(\sigma_{\phi\, \rm stat}^2 + 
\sigma_{\phi\,\rm orb}^2)^{1/2}$.  

In order to derive the differential correction to the spin frequency, 
$\Delta \nu_0$, and its derivative, $\dot \nu_0$, at the time $T_0$ we have to
derive a functional form for the time dependence of the phase delays.  We
started from the simple assumptions briefly summarized below: 
i) The bolometric luminosity $L$ is a good tracer of the mass 
accretion rate $\dot M$ {\it via} the relation $L = \zeta 
(GM/R) \dot M$, where $\zeta \leq 1$, and $G$, $M$, and $R$ are the 
gravitational constant and the neutron star mass and radius, respectively.
ii) The matter accretes through a Keplerian disk truncated at the 
magnetospheric radius, $R_{\rm m} \propto \dot M^{-\alpha}$, by its 
interaction with the (dipolar) magnetic field of the neutron star.  
At $R_{\rm m}$ the matter is forced to corotate with the
magnetic field of the neutron star and is funneled (at least in part) towards
the rotating magnetic poles, thus causing the pulsed emission. For standard
disk accretion $\alpha =2/7$; note that this is very close to the upper limit
$\alpha=2.3/7$ derived from our data. This upper limit on $\alpha$ can be
derived noting that the pulsations were clearly seen at the beginning of the 
observations in G05 at MJD$\sim 53342.0$, and were detected for 
$\Delta t \sim 11.5$ days until MJD$\sim 53353.5$, when the flux reduced by 
a factor of $\sim 10$.  The corresponding expansion in $R_{\rm m}$ must be 
$R_{\rm MAX}/R_{\rm MIN} = (1 - \Delta t/t_B^*)^{-\alpha} \leq R_{\rm CO}/R $ 
which, for $R = 10^6$ cm and $m=1.4$, gives 
$\alpha \leq \alpha_{\rm MAX} = 0.328 \simeq 2.3 /7 $ 
where $t_B^* \sim 12.4$ days is the decay time from G05 and $R_{\rm CO}$ is the 
corotation radius, both defined below.
Therefore we considered two extreme cases, namely $\alpha = 2/7$ and 
$\alpha = 0$, since a location of $R_{\rm m}$ independent of $\dot M$ has 
been proposed (see, {\it e.g.}, Rappaport, Fregeau, and Spruit, 2004).
iii) The matter accretes onto the neutron star its
specific Keplerian angular momentum at $R_{\rm m}$, $\ell = (GMR_{\rm
m})^{1/2}$, thus causing a material torque $\tau_{\dot M} = \ell
\times \dot M$. A firm upper limit to this torque is given by the condition
$\tau_{\dot M} \leq \ell_{\rm max} \times \dot M$, with $\ell_{\rm max} =
(GMR_{\rm CO})^{1/2}$, where $R_{\rm CO} = 1.50 \times 10^8 \; m^{1/3}
\nu^{-2/3}$ is the corotation radius (namely the radius at which the Keplerian
frequency equals $\nu$ and beyond which accretion is centrifugally inhibited),
and $m = M/{\rm M_\odot}$. 
iv) We {\it do not consider} any form of threading of the accretion disk by 
the magnetic field of the neutron star (see e.g.\ Ghosh \& Lamb 1979; Wang 1997; 
Wang 1996; Rappaport, Fregeau, and Spruit 2004 for a description of the magnetic 
threading), which implies that the only torque acting during accretion is 
$\tau_{\dot M}$. 

Under these hypotheses the spin frequency derivative is $\dot \nu = \ell \, 
\dot M/(2\pi \, I) $,
where $I$ is the moment of inertia of the neutron star and we 
have neglected any variation of $I$ caused by accretion. 
If $\dot M = \dot M(t)$, we have $\dot \nu(t) = (2\pi
\,I)^{-1} \, \ell_0 \, \dot M_0 \, (\dot M(t)/\dot M_0)^{1-\alpha/2}$, 
where $\ell_0 = (GMR_{\rm m \, 0})^{1/2}$, and
$R_{\rm m \,0}$ and $\dot M_0$ are $R_{\rm m}$ and $\dot M$ at $t = T_0$,
respectively. For the $\alpha = 0$ case we assumed $\ell_0 = \ell_{\rm max}$.
In this case we therefore assume that the system is accreting the maximum specific
angular momentum possible, giving an upper limit on the spin-up torque and
therefore a lower limit to the accretion rate $\dot M_0$.

Since we assumed $\dot M(t) \propto L(t)$, to determine the temporal
dependence of $\dot M(t)$ we studied 
the energy spectra of the source for each continuous interval of
data combining PCA and HEXTE data. All the spectra are well fitted with a
model consisting of a power law with an exponential cutoff plus thermal
emission from a Keplerian accretion disk modified by photoelectric absorption
and a Gaussian iron line. In order to derive $L(t)$ for each spectrum
we made the simple assumption $L(t) \propto F_{(3-150)}(t)$, which is
the unabsorbed flux in the RXTE PCA plus HEXTE energy band $(3-150)$ keV.
A good fit of $F_{(3-150)}(t)$ {\it vs} $t$ between December 7 and 14 
($\Delta t_{\rm obs} \sim 7$ days) is given by the expression 
$F_{(3-150)}(t) = F_{(3-150)} \, [1- (t-T_0)/t_B]$ with 
$t_B=8.4 \pm 0.1 $ days, where $F_{(3-150)}$ is the unabsorbed flux
at $t=T_0$.
Therefore we have $\dot \nu(t) = \dot \nu_0 \, [1-(t-T_0)/t_B]^{1-\alpha/2}$, 
where the spin frequency derivative at $t=T_0$ is
$\dot \nu_0 = (2\pi \,I)^{-1} \, \ell_0 \, \dot M_0$.

With this expression for $\nu(t)$ eq.~(\ref{eq:dotnudel}) can be readily
integrated. Since $\epsilon = (t-T_0)/t_B<1$ for $t<7$ days, we took a 
series expansion of the integral and obtained
$\delta \phi_{\dot \nu}(t) = 1/2 \, 
\dot \nu_0 \, (t-T_0)^2 \, [1-(2-\alpha)(t-T_0)/(6t_B) + \xi]$, with an 
error $\xi < \alpha(1-\alpha/2)/24 \, \epsilon^2$.

\begin{figure}
\plotone{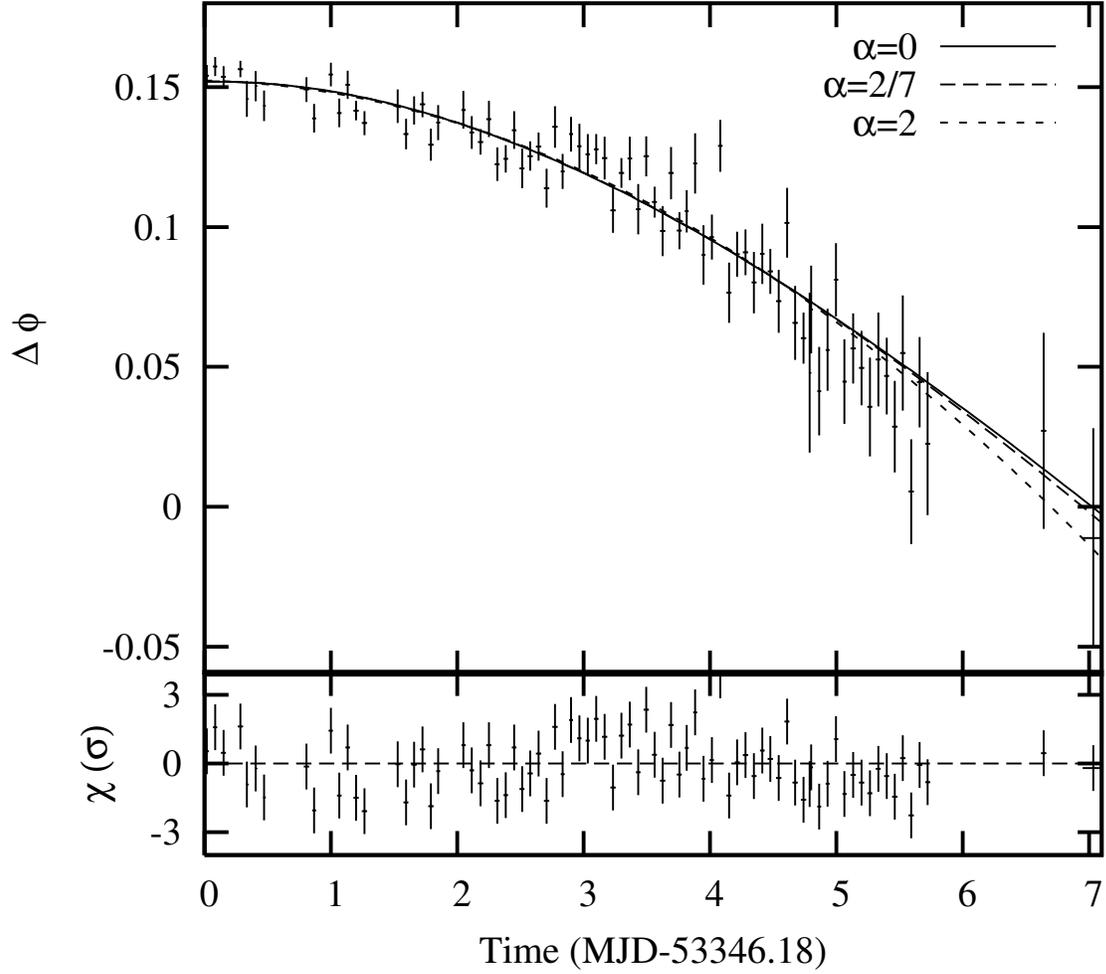}
\caption{Pulse phases computed folding at the spin period reported in 
Table~\ref{table1} and plotted versus time together with the best fit 
curves (upper panel) and residuals in units of $\sigma$ with respect to
the model with $\alpha = 2/7$ (lower panel). 
\label{fig1}}
\end{figure}

We have therefore fitted these phases with the function:
\begin{eqnarray}
\phi = - \phi_0 - \Delta \nu_0\, (t - T_0) - 
\frac{1}{2} \dot \nu_{0} (t - T_0)^2 
\left[ 1 - \frac{(2-\alpha)(t-T_0)}{6 t_B} \right] .
\label{eq:phi}
\end{eqnarray}
Using the best fit value for $\Delta \nu_0$ we computed the improved spin
frequency estimate and repeated the same procedure described at the beginning
of this paragraph, folding at the new estimate of the spin period. The new
phases were fitted with eq.~(\ref{eq:phi}). In this case, $\Delta \nu_0$ was
fully compatible with zero. These phases are plotted versus time in
Figure~\ref{fig1} (upper panel) together with the residuals in units of
$\sigma$ with respect to eq.~(\ref{eq:phi}) (lower panel). The best fit
estimates of $\nu_0$ and $\dot \nu_0$ are reported in Table~\ref{table1}
for three values of $\alpha$, namely $\alpha = 0$ which correspond
to a location of $R_{\rm m}$ independent of the accretion rate
(cfr. the model of Rappaport Fregeau, and Spruit 2004 in which $R_{\rm m}
= R_{\rm CO}$ for any $\dot M$), the standard case $\alpha = 2/7$ which
corresponds to $R_{\rm m}$ proportional to the Alfv\'en radius, 
and $\alpha = 2$
which has been given for comparison purposes and corresponds to a parabolic 
trend, expected in the case of constant $\dot M$. 
Of course, the value of $\dot \nu$ obtained in this latter case is in 
agreement with the value obtained by F05 of 
$\dot \nu = 8.4 (6) \times 10^{-13}$ Hz s$^{-1}$.
A comparison of the $\chi^2/$dof for each of the adopted values of 
$\alpha$ (also reported in Table~\ref{table1}) shows that the statistics 
is not good enough to distinguish between these three possibilities.  

Finally, to evaluate the systematic errors on the spin frequency and
its derivative, we adopted the positional uncertainty of $0.06''$ 
radius reported by Rupen et al. (2004) in a series expansion of 
eq.~(\ref{eq:posdel2}), finding $\sigma_{\nu\, {\rm syst}} \sim 2.2
\times 10^{-8}$ Hz and $\sigma_{\dot \nu\, {\rm syst}} \sim 4.4
\times 10^{-15}$ Hz/s, respectively. Even adopting a positional 
error circle of $0.2''$ (which is the distance between the 
optical and radio position, Fox \& Kulkarni 2004), the resulting
systematic uncertainties are $\sigma_{\nu\, {\rm syst}} \sim 7.3
\times 10^{-8}$ Hz and $\sigma_{\dot \nu\, {\rm syst}} \sim 1.5
\times 10^{-14}$ Hz/s. Note that the systematic error on the spin 
frequency is comparable with the error derived from the fit of the
phase delays and reported in Table~1, the systematic error on the spin 
frequency derivative is at least one order of magnitude below the
error derived from the fit.

\section{Discussion}

From the best-fit value of the spin frequency derivative $\dot \nu_0$
we can compute the mass accretion rate at $t = T_0$ through the
formula: 
\begin{equation}
\dot M_{-10} = 5.9 \times \dot \nu_{-13}\, I_{45}\,
m^{-2/3} (R_{\rm CO}/R_{\rm m \, 0})^{1/2},
\label{dotm0}
\end{equation}
where $\dot M_{-10}$ is $\dot M_0$ in units of $10^{-10}\, M_\odot$ yr$^{-1}$,
$\dot \nu_{-13}$ is $\dot \nu_0$ in units of $10^{-13}$ s$^{-2}$, and
$I_{45}$ is $I$ in units of $10^{45}$ g cm$^2$. In the following we will
adopt the FPS equation of state for the neutron star matter for $m = 1.4$
and the spin frequency of \igrj\ which gives $I_{45}= 1.29$ and 
$R= 1.14 \times 10^6$ cm (see e.g.\ Cook, Shapiro \& Teukolsky 1994).
This gives a lower limit in the mass accretion rate of $\dot M_{-10} 
\sim 70 \pm 10$ (case $\alpha = 0$).
In order to compare the experimental estimate of $\dot M_0$ with the
observed X-ray luminosity, we have to derive the bolometric luminosity
$L(t)$ from the observed flux $F_{(3-150)}(t)$. 
To this end we consider the spectral shape at $t=T_0$
in more detail. 

Since the value of the hydrogen column $N_{\rm H}$ is poorly
constrained in the RXTE energy band which starts at 2.5 keV, we fixed
it to $N_{\rm H} = 2.80 \times 10^{21}$ cm$^{-2}$, which is the
value obtained by Nowak et al.\ (2004) analysing Chandra data.
Our spectral results are practically independent 
of the precise value of the $N_{\rm H}$ below the total
Galactic hydrogen column in the direction of \igrj. 

The power law is the dominant spectral component. 
In particular we found a power law spectral index
$\alpha = -0.59^{+0.047}_{-0.034}$ and an $e$-folding energy 
$E_{\rm fold} \sim 178^{+83}_{-47}$ keV
(with values ranging from $60$ to over $260$ keV throughout our data). 
The ratio of the unabsorbed fluxes in the bands
$0.001-1000$ keV and $3-150$ keV is $1.6$. This ratio is almost independent 
on the $e$-folding energy, increasing up to $8\%$ when the $e$-folding 
energy increases from 60 to 260 keV. Therefore we assume
$F_{{\rm PL} \; (0,\infty)} \simeq 1.6 \times F_{{\rm PL} \; (3-150)}$.
The power-law component presumably originates in regions of small
optical depth just above each polar cap (see e.g.\ Poutanen \& Gierlinski 2003;
Gierlinski \& Poutanen 2005), thus we neglect, to first order, any effect
of the inclination of the emitting region with respect to the observer. 
On the other hand, we observe a single-peaked pulse profile, which means
that we only see the emission from one of these regions (e.g.\ Kulkarni \& 
Romanova 2005). 

If this is the case, we have to take into account the possibility that we
are underestimating the total flux in the power-law component because one
of the two polar caps is never visible 
(this can happen if the sum of the angles between the magnetic axis and spin 
axis and between the line of sight and the spin axis is less than $\pi/2$).
If we only see polar cap A, we indicate with $F_A$ the flux 
emitted by A reaching the observer. Because the emission from A is isotropic, 
to obtain the luminosity we have to integrate over $4 \pi d^2$, so we 
obtain $L_A = F_A 4 \pi d^2$. However, under our assumptions, this is an 
underestimate of the total (A + B) luminosity of the system consisting of the 
two polar caps, since part of the flux ($F_B$) that should be emitted in the 
direction of the observer is never visible.

Actually almost half of the emission from cap A facing the neutron star surface 
is intercepted by the neutron star itself and re-emitted towards the observer, 
and this re-emission should indeed be consider. 
However, it is reasonable to assume that this intercepted emission is reprocessed 
by the neutron star and re-emitted as a blackbody-like spectrum at a relatively 
low temperature. This reprocessed emission has therefore a very different 
spectral shape from the original power-law component: simple estimates demonstrate 
that the temperature associated with this blackbody-like emission is below 1 keV, 
and thus not related to the power-law component. Moreover, most of the emission 
from this component is outside the energy range of RXTE/PCA (and thus poorly 
constrained by the RXTE observation). 

Since the emission from an optically thin region is proportional to the volume 
of the emitting region visible, in the hypothesis that we totally miss the flux
from one of the two polar caps, we should multiply by a factor 2 
the unabsorbed flux of the power law in order to take into account the emission 
of the optically thin region above the unseen polar cap.
We therefore parametrize the luminosity of the power-law component with a factor
$\eta$ (which can assume values between 1 and 2), and we can write the total 
luminosity of the two polar caps as:
$L_{{\rm PL}} \simeq \eta F_{{\rm PL} \; (0,\infty)}
\times 4 \pi d^2 = 0.75^{+0.20}_{-0.15} \times 10^{37} \; \eta \; d_{\rm 5\,kpc}^2$ erg/s, 
where $d_{\rm 5\, kpc}$ is the source distance in units of $5$ kpc. 
The uncertainty on the luminosity has been evaluated conservatively by propagating 
the uncertainties on the spectral parameters treated as they were independent 
of each other. 

The second component is the thermal emission interpreted as emission from a 
Shakura-Sunyaev accretion disk, that is fitted with the {\tt diskbb} model. 
We found a temperature of the inner rim of the accretion disc of 
$T_{\rm in} = 0.68^{+0.20}_{-0.25}$ keV. On the other hand, because of the
poor coverage of RXTE/PCA at soft X-rays, the disk blackbody normalization,
$K = (R_{\rm in \, km}/D_{\rm 10 \, kpc})^2 \cos \theta = 25^{+170}_{-20}$ 
(where $D_{\rm 10 \, kpc}$ is the distance in units of 10 kpc),
is basically unconstrained. We therefore use the inner disk temperature and 
the Virial theorem to infer the bolometric luminosity of the disk, as follows.
We interpret the inner temperature of the accretion disk as derived
by the {\tt diskbb} model as the maximum temperature in the disk.
To prove this we have also fitted the soft X-ray spectrum with the 
{\tt diskpn} model (instead of the {\tt diskbb} model): {\tt diskpn} takes 
into account corrections for temperature distribution near the compact 
object using the Paczynski-Wiita pseudo-Newtonian potential 
(see Gierlinski et al. 1999). Also
this model gives a maximum temperature of the disk of 0.70 keV (fully
compatible with the value obtained with the {\tt diskbb} model). 
Standard disc theory (see e.g.\ Frank, King, \& Reine 2002) predicts
that the temperature of the disk attains a maximum value 
at a radius of $49/36$ $R_*$, where $R_*$ is the radius at which the disk is 
truncated, corresponding to $R_{\rm m \, 0}$ in our case. Using this in eq.~5.43
of Frank, King, \& Reine (2002) we obtain the inner disc radius through 
the relation: $R_{\rm m \, 0} = 1.77\, m^{1/3} \, 
\dot M_{-10}^{1/3} \, T_{\rm keV}^{-4/3}$ km. Combining this with 
eq.~\ref{dotm0}, and adopting $T_{\rm keV} = 0.70$, 
we solve for the mass accretion rate and the inner disc radius.
We found a mass accretion rate $\dot M_{-10} = 85 \pm 19$
and an inner disc radius $R_{\rm m \, 0} \simeq 
1.46^{+0.62}_{-0.49} \times 10^6$ cm, that this is exactly 
in the very narrow range between the neutron star radius ($\sim 10^6$ cm) and the 
corotation radius ($\sim 2.4 \times 10^6$ cm); the agreement with 
the expectation is compelling. The Virial theorem allows to calculate the 
fraction of the total luminosity that is emitted by the disc: 
$0.5 R/R_{\rm m\,0} = 0.39$. Therefore 
$L_{{\rm BB\; 0}} = 0.39 / (1-0.39) \times L_{{\rm PL} \; 0}
= 4.8 \times 10^{36} \, \eta \, d_{\rm 5\,kpc}^2$ erg/s. 

The total bolometric luminosity is therefore $L_0 = 1.23^{+0.45}_{-0.15} 
\times 10^{37} \, \eta \, d_{\rm 5\,kpc}^2$ erg/s. 
If we compare this luminosity with the mass accretion rate inferred from the
timing analysis (assuming an efficiency $\zeta = 1$) we obtain a distance to
the source in the range $(10.5 - 15) / \eta^{1/2}$ kpc. Note that 10 kpc is close 
to the edge of our Galaxy in the direction of \igrj.
If we push the factor $\eta$ to its maximum value of 2, we obtain a more reasonable 
range of distances to the source of $7.4 - 10.7$ kpc, consistent with the lower 
limit of 5 kpc discussed in F05.  
Note also that the effect of including magnetic torques due to threading (see 
assumption iv in \S~3) can only push the source further away. 
This is evident from the $\alpha = 0$ case (which gives a spin derivative similar 
to the $\alpha = 2/7$ case discussed here), in which we assume that the system is 
accreting the specific angular momentum at the corotation radius, that is the 
maximum possible; since any torque beyond the corotation could only spin down 
the system, this would increase the required $\dot M$ to justify the measured 
spin-up. 

Finally, using the value of the inner disc radius derived above and the normalization 
of the disc blackbody model, $K = (R_{\rm in \, km}/D_{\rm 10 \, kpc})^2 \cos \theta 
= 25^{+170}_{-20}$, 
we infer the inclination of the system with respect to the line of sight, 
which is $i \ge 40^\circ$
for a distance of 9.5 kpc. 

\section{Conclusions}

We have analysed RXTE data of the fastest known accreting millisecond 
pulsar, \igrj, during the period $7 - 14$ December, 2004. We report a revised 
estimate of the spin period and discuss the spin period derivative. The source
shows a strong spin-up, which indicates a mass accretion rate of about
$8.5 \times 10^{-9} \, M_\odot$ yr$^{-1}$. 
We have checked that this mass accretion rate
is compatible with the X-ray spectrum of the source. In particular we have shown
that, with this high accretion rate, the measured temperature of the disc
blackbody emission implies an inner disc radius in excellent agreement with  
the inferred magnetospheric radius of the source (constrained to be in the 
narrow range $(1 - 2) \times 10^6$ cm) and that the source is probably 
seen at high inclination. 
However, the mass accretion rate inferred by the observed spin-up (and calculated
using standard values for the neutron star moment of inertia), using the 
simple relation $L_{\rm bol} = \zeta G M \dot M / R$ (with an efficiency 
$\zeta = 1$), would correspond to a quite high bolometric luminosity of the 
source of $L_{\rm bol} \sim 10^{38}$ ergs/s, much higher than the observed
source luminosity assuming a distance of 5 kpc. We propose that the simplest
explanation of this discrepancy is that part of the accretion luminosity 
is not visible. Indeed, if we only see the emission of one of the two polar
caps, we could miss up to half of the flux in the power-law component. Under
this hypothesis (described in detail in the previous section), we have extrapolated 
the flux of the source derived from the X-ray spectrum, which corresponds to
a bolometric luminosity of $L_{\rm bol} \simeq 2.5 \times 10^{37} \, 
d_{\rm 5\,kpc}^2$ erg/s. 
Comparing this extrapolated bolometric luminosity with the mass accretion rate of 
the source as derived from the timing, we find an agreement between these two 
quantities if we place the source at a distance between 7 and 10 kpc. 

Other possibilities to explain this discrepancy can be that part of the 
accretion luminosity is not observed because emitted in other energy bands or 
because the efficiency $\zeta$ of the conversion of the gravitational potential
energy of the accreting matter into X-ray luminosity is less than $1$, or because 
of occultation effects (which may be favoured if indeed the source is highly 
inclined). In these cases we should conclude that the observed X-ray luminosity 
is not a good tracer of the total mass accretion rate, $\dot M$, onto the neutron star.

\acknowledgements
This work was partially supported by the Ministero della Istruzione, 
della Universit\`a e della Ricerca (MIUR).



\begin{deluxetable}{llll}
\tabletypesize{\scriptsize}
\tablecaption{Orbital and spin parameters of \igrj. \label{tab1}}
\tablehead{
\colhead{} &
\colhead{G05} &
\colhead{F05} &
\colhead{This work}
}
\startdata
Projected semimajor axis, $\asini$ (lt-ms)                      & $64.993(2)$ &   --  & --  \\
Orbital period, $\Porb$ (s)                                     & $8844.092(6)$ & --  & -- \\
Epoch of ascending node passage, ${T^*}$\tablenotemark{a} (MJD) & $53345.1619258(4)$  & --  & -- \\
Eccentricity, $e$	                                        & $<2 \times 10^{-4}$ (3 $\sigma$)  
& -- & --   \\
Spin frequency, $\nu_0$ (Hz)                                    & $598.89213064(1)$ 
& $598.89213060(1)$ & $598.89213053(2)$ \\
Spin frequency derivative, $\dot \nu_0$ (Hz/s) ($\dot \nu =$ constant) &  
$< 8 \times 10^{-13} $ (3 $\sigma$)  & $8.4(6) \times 10^{-13}$ &
$0.85(0.11)\times 10^{-12} \; \; $  ($\chi^2/$dof=106/77 ) \\  
Spin frequency derivative, $\dot \nu_0$ (Hz/s) ($\alpha=0$)\tablenotemark{b}  &  -- 
& -- & $1.17(0.16)\times 10^{-12} \; \;$  ($\chi^2/$dof=113/77 )  \\  
Spin frequency derivative, $\dot \nu_0$ (Hz/s) ($\alpha=2/7$)\tablenotemark{b} &  -- 
& -- & $1.11(0.16)\times 10^{-12} \; \; $  ($\chi^2/$dof=111/77 ) \\  
Epoch of the spin period, ${T_0} $  (MJD)    &  --  & 53346.0   &  53346.184635
\enddata
\tablecomments{Errors are given at $1 \sigma$ confidence level. The
errors quoted for the spin frequency and spin frequency derivative
are derived from the phase delays fitting, and do not
include systematic errors induced by the source position
uncertainty. Adopting a positional uncertainty of $0.06''$ (Rupen 
et al.\ 2004), these are: $\sigma_{\nu\, {\rm syst}} \sim 2.2
\times 10^{-8}$ Hz and $\sigma_{\dot \nu\, {\rm syst}} \sim 4.4
\times 10^{-15}$ Hz/s, respectively.}

\tablenotetext{a}{G05 reported a value of $53345.1875164(4)$ MJD
for the epoch of superior conjunction, i.e. when the NS is behind the
companion; as, in this work, we considered the epoch of ascending node passage
as a reference time, the G05 reference time reported here has been
decremented by $\Porb/4$.}
\tablenotetext{b}{Averaging $\dot \nu_0$ over the 7 days of our observation, we
get $<\dot \nu_0> \simeq 0.68 \times 10^{-12}$ Hz/s.}
\label{table1}
\end{deluxetable}

\clearpage

\end{document}